\documentclass[aps, prl, twocolumn,10pt,showpacs,superscriptaddress,amsmath,amssymb,nobibnotes]{revtex4-1}
\usepackage{graphicx}
\usepackage{amsfonts}    
\usepackage{verbatim}   
\usepackage{xcolor}      
\usepackage{subfigure}  
\usepackage{booktabs}
\usepackage{threeparttable}
\usepackage{dcolumn}
\usepackage{multirow}
\usepackage{physics}
\usepackage{amsmath}

\usepackage{dsfont}
\usepackage{enumerate}
\usepackage{comment}
\usepackage[final,colorlinks,linkcolor=blue,anchorcolor=blue,urlcolor=blue,citecolor=blue]{hyperref}

\begin{document}

\title{ Experimental demonstration of fully passive quantum key distribution  }

\author{Feng-Yu Lu}\email{These authors contributed equally to this work}
\author{Ze-Hao Wang}\email{These authors contributed equally to this work}
\affiliation{CAS Key Laboratory of Quantum Information, University of Science and Technology of China, Hefei, Anhui 230026, P. R. China}
\affiliation{CAS Center for Excellence in Quantum Information and Quantum Physics, University of Science and Technology of China, Hefei, Anhui 230026, P. R. China}
\author{Víctor Zapatero}\email{These authors contributed equally to this work}
\affiliation{Vigo Quantum Communication Center, University of Vigo, Vigo E-36310, Spain}
\affiliation{Escuela de Ingeniería de Telecomunicación, Department of Signal Theory and Communications, University of Vigo, Vigo E-36310, Spain}
\affiliation{AtlanTTic Research Center, University of Vigo, Vigo E-36310, Spain}
\author{Jia-Lin Chen}
\affiliation{CAS Key Laboratory of Quantum Information, University of Science and Technology of China, Hefei, Anhui 230026, P. R. China}
\affiliation{CAS Center for Excellence in Quantum Information and Quantum Physics, University of Science and Technology of China, Hefei, Anhui 230026, P. R. China}
\author{Shuang Wang}\email{wshuang@ustc.edu.cn}
\author{Zhen-Qiang Yin}\email{yinzq@ustc.edu.cn}
\affiliation{CAS Key Laboratory of Quantum Information, University of Science and Technology of China, Hefei, Anhui 230026, P. R. China}
\affiliation{CAS Center for Excellence in Quantum Information and Quantum Physics, University of Science and Technology of China, Hefei, Anhui 230026, P. R. China}
\affiliation{Hefei National Laboratory, University of Science and Technology of China, Hefei 230088, China}
\author{Marcos Curty}
\affiliation{Vigo Quantum Communication Center, University of Vigo, Vigo E-36310, Spain}
\affiliation{Escuela de Ingeniería de Telecomunicación, Department of Signal Theory and Communications, University of Vigo, Vigo E-36310, Spain}
\affiliation{AtlanTTic Research Center, University of Vigo, Vigo E-36310, Spain}
\author{De-Yong He}
\affiliation{CAS Key Laboratory of Quantum Information, University of Science and Technology of China, Hefei, Anhui 230026, P. R. China}
\affiliation{CAS Center for Excellence in Quantum Information and Quantum Physics, University of Science and Technology of China, Hefei, Anhui 230026, P. R. China}
\affiliation{Hefei National Laboratory, University of Science and Technology of China, Hefei 230088, China}
\author{Rong Wang}
\affiliation{Department of Physics, University of Hong Kong, Hong Kong SAR, China}
\author{Wei Chen}
\affiliation{CAS Key Laboratory of Quantum Information, University of Science and Technology of China, Hefei, Anhui 230026, P. R. China}
\affiliation{CAS Center for Excellence in Quantum Information and Quantum Physics, University of Science and Technology of China, Hefei, Anhui 230026, P. R. China}
\affiliation{Hefei National Laboratory, University of Science and Technology of China, Hefei 230088, China}
\author{Guan-Jie Fan-Yuan}
\affiliation{CAS Key Laboratory of Quantum Information, University of Science and Technology of China, Hefei, Anhui 230026, P. R. China}
\affiliation{CAS Center for Excellence in Quantum Information and Quantum Physics, University of Science and Technology of China, Hefei, Anhui 230026, P. R. China}
\author{Guang-Can Guo}
\author{Zheng-Fu Han}
\affiliation{CAS Key Laboratory of Quantum Information, University of Science and Technology of China, Hefei, Anhui 230026, P. R. China}
\affiliation{CAS Center for Excellence in Quantum Information and Quantum Physics, University of Science and Technology of China, Hefei, Anhui 230026, P. R. China}
\affiliation{Hefei National Laboratory, University of Science and Technology of China, Hefei 230088, China}

\date{\today}

\begin{abstract}
The passive approach to quantum key distribution (QKD) consists of removing all active modulation from the users' devices, a highly desirable countermeasure to get rid of modulator side-channels. Nevertheless, active modulation has not been completely removed in QKD systems so far, due to both theoretical and practical limitations. In this work, we present a fully passive time-bin encoding QKD system and report on the successful implementation of a modulator-free QKD link. According to the latest theoretical analysis, our prototype is capable of delivering competitive secret key rates in the finite key regime.

\end{abstract}

\maketitle


Quantum key distribution (QKD)\cite{BB84} is one of the most successful applications of quantum information science, since it allows for information-theoretically secure communications between two distant users regardless of the ---potentially unlimited--- computational power of an eavesdropper Eve~\cite{lo1999unconditional,shor2000simple,scarani2009security,Rennersecurity}. However, information-theoretic security may be compromised by the presence of loopholes in the QKD equipment, which may open side-channels for Eve to obtain information in unexpected ways \cite{lutkenhaus2002quantum,qi2007time,makarov2009controlling,makarov2006effects,zhao2008quantum,lydersen2010hacking,jain2011device,sajeed2016insecurity,qian2018hacking,lutkenhaus2002quantum,liu2011proof,mailloux2016using,huang2019laser,tamaki2016decoy,pang2020hacking,huang2020laser,zhang2021securing,mizutani2019quantum,zapatero2021security,nagamatsu2016security,grunenfelder2020performance,yoshino2018quantum,roberts2018patterning,lu2021intensity}. 

Measurement-device-independent (MDI) QKD \cite{lo2012measurement,braunstein2012side,wang2015phase,tang2016experimental,wang2019practical,navarrete2021practical,lu2022unbalanced,lu2023hacking} has been proposed to close all detection-related security loopholes. Thus, in recent years, more and more attention is set on the source-side loopholes~\cite{lutkenhaus2002quantum,gisin2006trojan,lucamarini2015practical,tamaki2016decoy,yoshino2018quantum,huang2019laser,pang2020hacking,huang2020laser,zhang2021securing,lu2022unbalanced}. For instance, Eve can obtain some modulation information with a Trojan-Horse attack~\cite{gisin2006trojan,lucamarini2015practical}. Also, the injection of bright light or external magnetic fields may damage the source-side devices~\cite{huang2019laser,pang2020hacking,huang2020laser,tan2022external}. Similarly, correlations in modulations may invalidate the independent and identically distributed assumption, opening the door to sophisticated attacks~\cite{pereira2020quantum,zapatero2021security,sixto2022security,yoshino2018quantum,roberts2018patterning,lu2021intensity}.

In this context, some passive schemes have been proposed to deal with source side-channels. For instance, by making use of the photon-number relationship \cite{curty2009non,curty2010passiveSources,adachi2007simple,ma2008quantum,mauerer2007quantum,krapick2014bright,wang2016scheme} between two pulses, Alice can deduce the photon number distribution of one of them by detecting the other, in so passively implementing the decoy-state technique~\cite{hwang2003quantum,wang2005beating,lo2005decoy,curty2010passiveDecoy,adachi2007simple}. Besides, superimposing two mutually orthogonal polarizations in a polarizing beam splitter (BS) and further incorporating a post-selection method, one can passively encode the four BB84 states.

However, the above passive schemes can only replace part of the active modulations, and therefore they only provide partial solutions. In spite of the considerable efforts made, passively implementing both the decoy-state choice and the BB84 encoding with linear optical components and laser sources has been a pending task for a decade. To state it shortly, the basic difficulty is dealing with the fact that the passive decoy-states and the passive encoding states are correlated. Moreover, it is also a challenge to prepare and post-select the required decoy-states and encoding states simultaneously, due to the large difference between passive QKD systems and traditional active QKD systems. Recently, two solutions against this elusive problem have been proposed \cite{wang2023fully,zapatero2023fully}, finally paving the way towards fully passive (FP) QKD.

In this Letter, we propose a FP time-bin encoding QKD source and report on the first successful implementation of a FP-QKD link. Our prototype can efficiently prepare different decoy-states and encoding states simultaneously, and our post-selection module accurately post-selects the required states and ignores the undesired rounds to reduce the throughput and computation. As a result, it is capable of delivering competitive secret key rate (SKRs) in the finite key regime, following the security analysis in~\cite{Zapatero2023finitekey}. What is more, compared with the multi-laser approach introduced in~\cite{wang2023fully}, the single-laser structure we deploy allows to avoid wavelength side-channels. Putting it all together, the scheme is of great significance to reach higher implementation security and promote the applicability of QKD.

\begin{figure*}[htbp]
	\includegraphics[width=13cm]{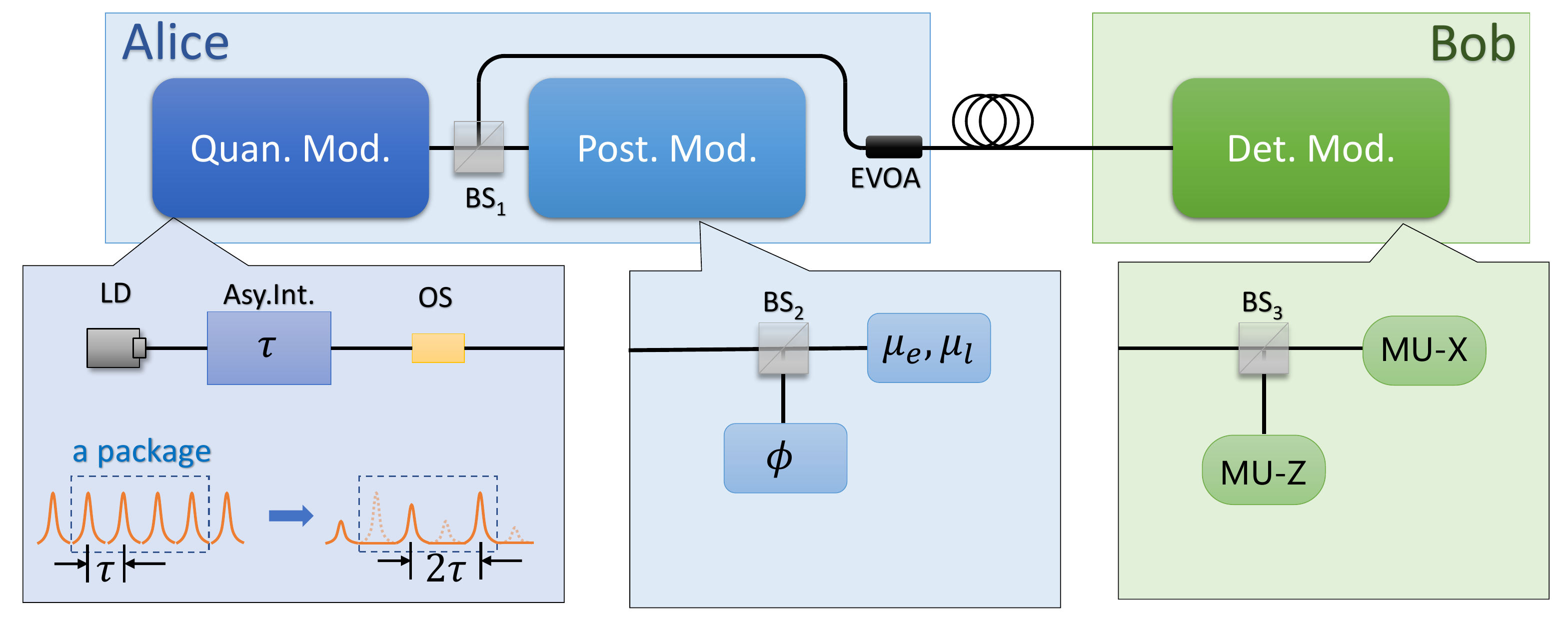}
	\caption{\label{fig: fully passive scheme} Schematic diagram of our FP-QKD scheme. Quan. Mod.: quantum module; Post. Mod.: post-selection module; Det. Mod.: detection module; LD: laser diode; Asy. Int.: asymmetric interferometer.; OS: optical switch; BS: beam splitter; EVOA: electronic variable optical attenuator; MU-Z(X): measurement unit for measuring the $Z$ ($X$) basis. $\tau$: length difference between the two paths; $\mu_{e(l)}$: measurement of the intensity of the early (late) time-bin; and $\phi$: measurement of the relative phase between the two time-bins.} 
\end{figure*}

As illustrated in Fig.~\ref{fig: fully passive scheme}, our FP-QKD setup consists of three modules. In the source side, the quantum module is employed to passively and randomly generate different quantum states and intensities. Then, the post-selection module locally measures the prepared states and intensities to post-select the bases, raw key bits, and decoy-states. The third module is Bob's detection module.

In the quantum module, a LD generates a train of phase-randomized coherent states with a certain period, say $\tau$. In order to obtain four degrees of freedom~\cite{wang2023fully}, we treat every four consecutive pulses as a ``package'' 
\begin{equation}
	\begin{aligned}
		\ket{\sqrt{\mu_{\rm{in}}}e^{i\phi_1} }_1 \ket{\sqrt{\mu_{\rm{in}}}e^{i\phi_2} }_2 \ket{\sqrt{\mu_{\rm{in}}}e^{i\phi_3} }_3 \ket{\sqrt{\mu_{\rm{in}}}e^{i\phi_4} }_4
	\end{aligned}
\end{equation}
 where $\mu_{\rm{in}}$ is the output intensity of the LD, and $\phi_1$ to $\phi_4$ are the phases of the four pulses. The pulse train is then fed to an asymmetric interferometer whose path difference is $\tau$, such that adjacent pulses interfere at the output port of the interferometer. After that, an OS opens and closes with fixed period $2\tau$ to eliminate the odd pulses, which includes the interference of $\ket{\sqrt{\mu_{\rm{in}}}e^{i\phi_2} }_2$ and $\ket{\sqrt{\mu_{\rm{in}}}e^{i\phi_3} }_3$ in every package, and the interference between pulses belonging to different packages. The remaining two pulses in a package constitute the late and the early time-bins of an encoded state. Importantly as well, the operation of the OS is fixed and uncorrelated to the protocol settings, such that it does not constitute a side channel in this regard. Nevertheless, a practical OS with a finite extinction ratio may leak partial information about the neighboring pulses~\cite{wang2023fully}. A short discussion on this potential side channel is included in Supplemental Material III.
 
The output of the quantum module is unevenly split: most of the intensity is sent to the post-selection module for Alice's local measurement, and the remaining part is attenuated to the single-photon level and sent to Bob. This can be expressed as
\begin{equation}
\begin{aligned}
\label{eq: coherent state}
&\ket{\sqrt{\mu_{\rm{max}}}\left(e^{i\phi_1}+ e^{i\phi_2}\right)/2}_e \ket{\sqrt{\mu_{\rm{max}}}\left(e^{i\phi_3}+ e^{i\phi_4}\right)/2}_l\\
& = \ket{ \sqrt{\mu_e} e^{i\phi_e}  }_e \ket{ \sqrt{\mu_l} e^{i\phi_l}  }_l,
\end{aligned}
\end{equation}
where $\mu_{\rm{max}} = 2\mu_{\rm{in}}\eta_{S}$ stands for the maximum intensity in each time-bin, $\eta_S$ denoting the total attenuation of the source-side devices. Similarly, $\mu_e = \mu_{\rm{max}}\big[1 + \cos(\phi_1 - \phi_2) \big]/2$ and $\mu_l = \mu_{\rm{max}}\big[1  + \cos(\phi_3 - \phi_4) \big]/2$ denote the respective intensities of the early time-bin and the late time-bin, while $\phi_e = (\phi_1 + \phi_2)/2$ and $\phi_l = (\phi_3 + \phi_4)/2$ denote their phases. Hence, defining $\phi_G = \phi_e$, $\phi = \phi_l - \phi_e$, $\mu = \mu_e + \mu_l$ and $\theta = 2\arccos{\sqrt{\mu_e/\mu}}$ respectively as the global phase, the relative phase, the intensity and the polar angle of the time-bin state. The corresponding single-photon state can be expressed as $\cos(\theta/2) \ket{e} + e^{i\phi} \sin(\theta/2) \ket{l}$, where $\ket{e} = \ket{1}_e\otimes{}\ket{0}_l $ ($\ket{l} = \ket{0}_e\otimes{}\ket{1}_l$) denotes the early (late) time-bin single-photon state.

\begin{figure}[htbp]
	\includegraphics[width=8.8cm]{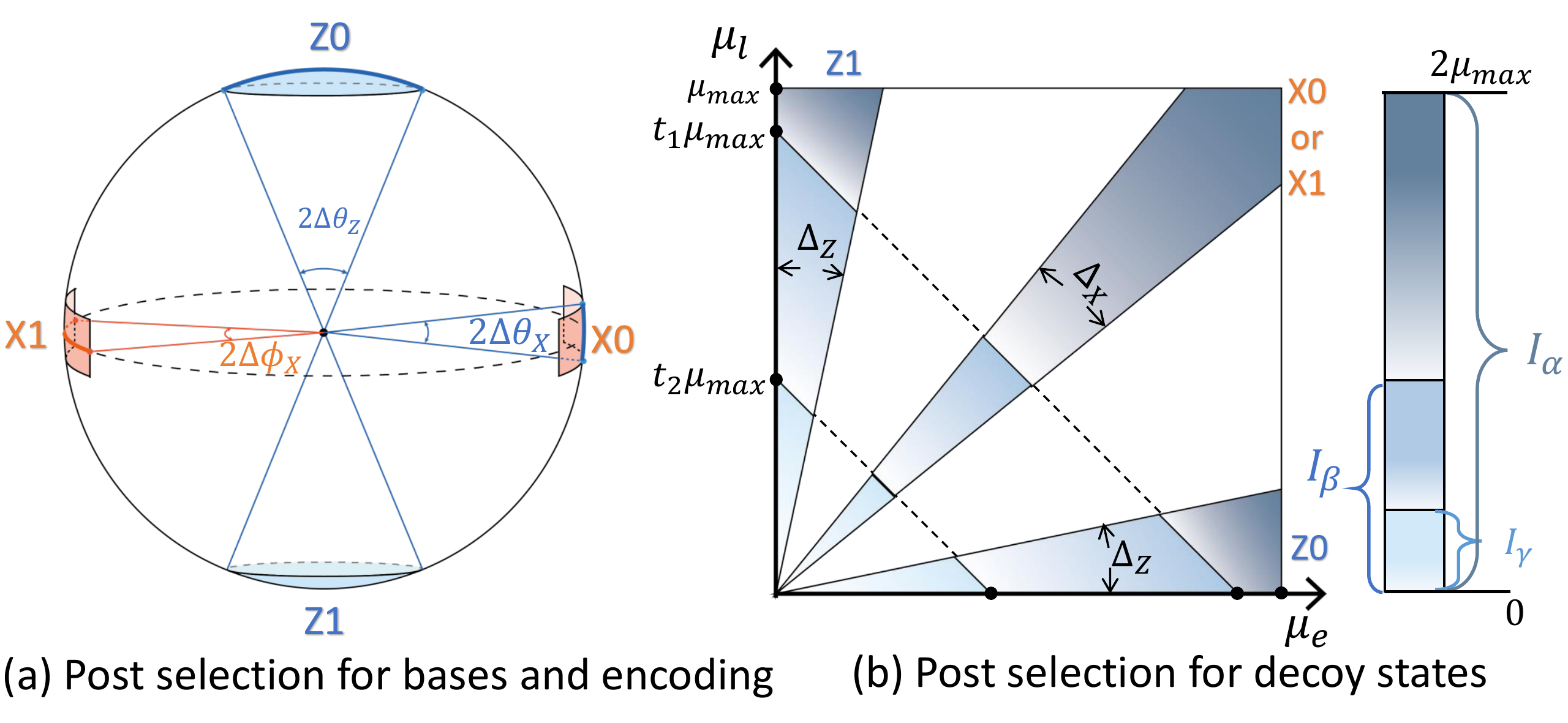}
	\caption{\label{fig: decoy-state selection} Schematic illustration of the post-selection. (a) Post-selection of bases and bit values. $\Delta\theta_Z$, $\Delta\theta_X$, and $\Delta\phi_X$ are pre-decided thresholds that characterize the acceptance regions. The areas Z0 and Z1 (X0 and X1) define the key(test)-basis, while the blank area corresponds to the rejected data. (b) Post-selection of decoy-states. The horizontal (vertical) axis denotes the intensity $\mu_{e(l)}$ of the early (late) time-bin, and the total intensity at any point of the graph is $\mu=\mu_e+\mu_l$. As stated in the main text, $\mu_{\rm{max}}$ denotes the maximum intensity in each time-bin. On the other hand, $\Delta_Z$ and $\Delta_X$ are threshold values related to $\Delta \theta_Z$, $\Delta\theta_X$ and $\Delta\phi_X$, and $t_1$ and $t_2$ are pre-decided thresholds that define the decoy-state intervals. Specifically, overlapping intensity intervals are used in the experiment (see Supplemental Material~I.}
\end{figure}

\begin{figure*}[htbp]
	\includegraphics[width=17cm]{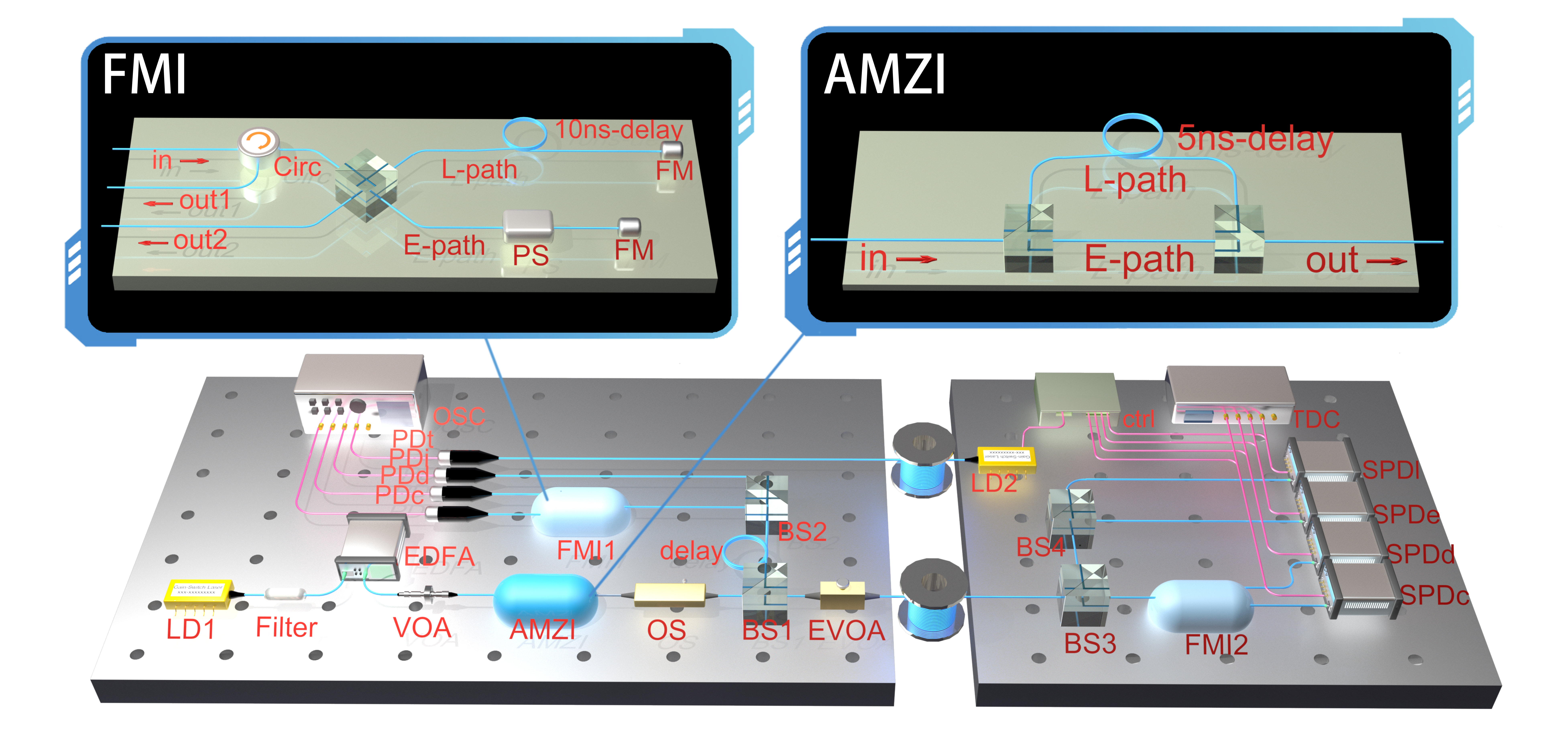}
	\caption{\label{fig: exp set-up} Depiction of the experimental setup. Devices in the main setup: LD: laser diode, VOA: variable optical attenuator, EDFA: erbium-doped fiber amplifier, AMZI: asymmetric Mach-Zehnder interferometer, FMI: Faraday-Michelson interferometer, OS: optical switch, BS: beam splitter, EVOA: electronic variable optical attenuator, SPD: single-photon detector, PD: photon diode, TDC: time-digital converter, OSC: oscilloscope, ctrl: computer and FPGA, Devices in FMI and AMZI: PS: phase shifter, FM: faraday mirror; Circ: circulator; L(E)-path: `late' (`early') path of the FMI and the AMZI. }
\end{figure*}
In the detection module, Bob passively selects the $Z$ or the $X$ basis to measure the received states, publicly announces if a successful detection event occurs in the measurement unit (MU), and records his raw key bits and bases according to the outcomes of the MUs. For those rounds where a successful event is announced, Alice records (or discards) her raw key bits, bases, and decoy settings according to the post-selection method depicted in Fig.~\ref{fig: decoy-state selection}. A complete description of the post-selection is presented in Supplemental Material~I.

Once the quantum communication ends, Alice and Bob reveal their basis choices for sifting, and disclose parts of their raw key data for the estimation of the secret key length. The details of the parameter estimation method are summarized in Supplemental Material~II, sticking to the finite-key analysis provided in~\cite{Zapatero2023finitekey}.

The experimental setup is illustrated in Fig.~\ref{fig: exp set-up}. A gain-switched LD driven by a homemade circuit generates pulse trains with 200 ps pulse width and 5 ns interval, and the phase of each pulse is randomized due to the amplified spontaneous emission process \cite{li2023high}. By carefully tuning the temperature-electronic control in the LD, the central-wavelength of the LD is locked at 1550.12 nm. An EDFA is employed to amplify the intensity, and a VOA is employed to protect Alice's local PDs. The pulses are fed to an asymmetric AMZI whose path difference is 5 ns to generate random intensities. A 5 GSa/s arbitrary waveform generator connected to a radio-frequency amplifier produces a 100 MHz square-wave periodic signal to drive a $\mathrm{LiNbO_3}$ OS to eliminate the odd pulses. The interval between the early and late time-bins is 10 ns and the interval between two time-bin states is 20 ns, meaning that our system works at 50 MHz period. The 1:99 BS $\mathrm{BS_1}$ sends $99\%$ of the intensity to the post-selection module for Alice's local measurement and, as mentioned above, the remaining $1\%$ is attenuated to a single-photon level by an EVOA with attenuation ratio $\eta_A$, to be ultimately sent to Bob through a fiber quantum channel.

At the detection side, Bob passively selects the $Z$ or the $X$ basis to measure the received time-bin states. The detection module consists of a BS ($\rm{BS_3}$) and two MUs. MU-Z is the unit for measuring the $Z$ basis, and it consists of a 50:50 BS ($\rm{BS_4}$) and two homemade SPDs. The two SPDs work on the gated mode \cite{he2017sine} and respectively open their gates at the early and the late time-bins. That is to say, Bob records bit 0 (1) in the $Z$ basis when $\rm{SPD_e}$ ($\rm{SPD_l}$) clicks. Analogously, MU-X measures the $X$ basis and consists of a FMI ($\rm{FMI_2}$) and two homemade SPDs ($\rm{SPD_c}$ and $\rm{SPD_d}$). The path difference of the $\rm{FMI_2}$ is 10 ns, such that the adjacent pulses interfere in its output port. $\rm{SPD_c}$ and $\rm{SPD_d}$ are connected to the two outputs of the $\rm{FMI_2}$ to measure constructive and destructive interference, respectively. These two SPDs also work on the gated mode, with a 50 MHz frequency to filter the dark counts and the interference between different packages. We observe a dark count rate of $6\times10^{-7}$ in the detection module, while its overall detection efficiency is $12.5\%$. As for the detection events, Bob records bit 0 (1) in the $X$ basis when $\rm{SPD_c}$ ($\rm{SPD_d}$) clicks. The output signals of each SPD are copied by a homemade circuit. Bob sends one of the copies to the time-digital converter for his raw key bit generation, and the other copy is sent to a homemade AND-logic circuit accompanied by a homemade electrical-to-optical converter. A click from any of the SPDs triggers the converter to generate a light pulse to be sent to Alice. When Alice receives this pulse, she measures and records the corresponding $\mu_e$, $\mu_l$, and $\phi$ using her post-selection module.

In Alice's post-selection module, the pulses are evenly split by $\mathrm{BS_2}$. The first part is measured by a 20-GHz bandwidth high-speed photon diode $\rm{PD}_i$ and the other part is fed to $\rm{FMI_1}$, whose path difference is 10 ns. The early and the late time-bin pulses interfere at the output port of $\rm{FMI_1}$, and the constructive and destructive interference results are respectively detected by two 20-GHz bandwidth high-speed PDs, $\rm{PD}_c$ and $\rm{PD}_d$. The outputs of the three PDs are monitored by a 20-GSa/s high-speed oscilloscope and stored if Bob announces a successful event, or discarded otherwise. If stored, Alice calculates the intensities $I_e$ and $I_l$ according to the measurement results of $\rm{PD}_i$ \cite{lu2021intensity}, and the intensities $I_c$ and $I_d$ according to the respective measurement results of $\rm{PD}_c$ and $\rm{PD}_d$, $I_e$ ($I_l$) denoting the intensity of the early (late) time-bin. Here, $I_c$ and $I_d$ stand for the constructive and the destructive interference results, respectively. The central parameters for Alice's post-selection are calculated as $\mu_{e(l)} = I_{e(l)} \eta_A t_{B_1}/\left(h\nu t_{B_2}(1 - t_{B_1})\right)$ and $\phi = \pi \pm \arccos\left(1 -{ 2I_c} / (I_c + I_d) \right)$,
where $t_{B_{1(2)}}$ is the transmittance of $\rm{BS_{1(2)}}$, $h$ is the Planck constant, and $\nu = 1550.12$ nm is the central wavelength. According to the post-selection rule, Alice deduces and records her basis, raw key bit, and decoy setting.

We successfully proved the feasibility of FP-QKD in three different scenarios: at 6 dB of channel loss with $N=10^{10}$ transmitted signals (experiment 1), at 10 dB of channel loss with $N=10^{10}$ transmitted signals (experiment 2), and at 6 dB of channel loss with $N=10^{9}$ transmitted signals (experiment 3). In all three of them, the post-selection thresholds ---presented in Fig.~\ref{fig: decoy-state selection}--- are set to $\Delta\theta_Z=0.6135$, $\Delta\theta_X=0.1002$, $\Delta\phi_X=0.6435$, $t_2=0.15$ and $t_1=0.55$. The test-basis probability of Bob's passive module is $q_X=0.25$, and the value of $\mu_{\rm{max}}$ is determined to be 0.359 in the first experiment and $0.254$ in the other two. Lastly, the finite key parameter introduced in Supplemental Material~II is set to $\epsilon=10^{-20}$ for illustration purposes, leading to an overall secrecy parameter of $\epsilon_{\rm sec}\approx{}4\times{}10^{-10}$ and a correctness parameter of $\epsilon_{\rm cor}=10^{-20}$.

\begin{figure}[htbp]
	\includegraphics[width=8cm]{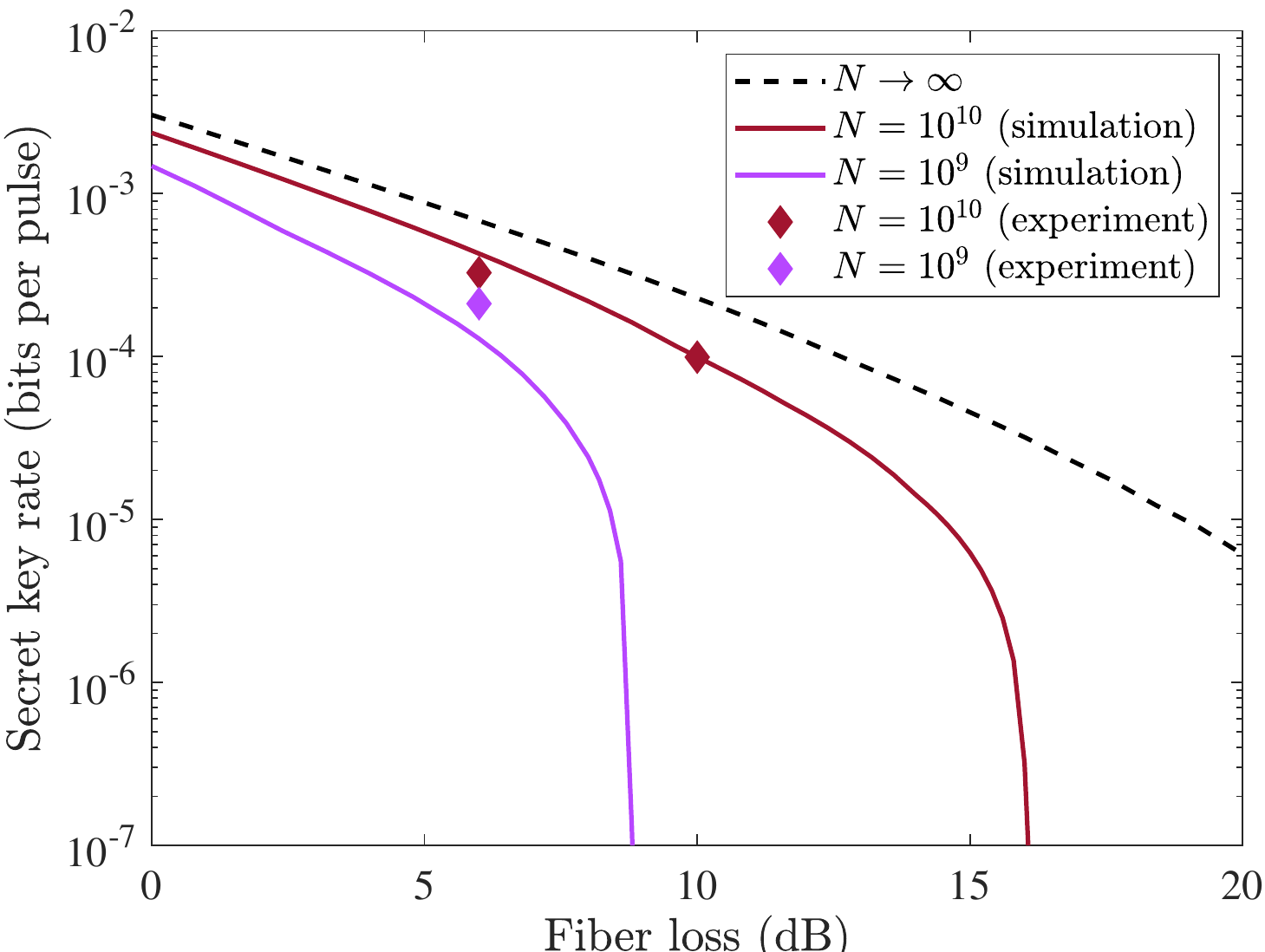}
    	\caption{\label{fig: exp SKR} Experimental and simulated performance of our FP-QKD system as a function of the fiber loss. The dashed line represents the simulated SKR in the asymptotic limit, while the solid lines represent the simulated SKRs with data sizes $N=10^{10}$ and $10^{9}$, as indicated in the legend. Lastly, the diamond-shaped points denote the experimental SKRs with data sizes $N=10^{10}$ and $10^{9}$, respectively. In the simulations, the value of $\mu_{\rm{max}}$ is optimized as a function of the channel loss, while all other settings are fixed as in the experiments.}
\end{figure}

As shown in Fig.~\ref{fig: exp SKR}, the extractable SKR is about $3.3\times{}10^{-4}$ ($9.9\times{}10^{-5}$) bits per pulse in experiment 1 (2), and $2.1\times{}10^{-4}$ in experiment 3. In addition, in Tab.~\ref{tab: experimental results} we show the sifted-key length ($M^{Z}_{\alpha}$), the lower bound on the number of key-basis single-photon counts ($M^{\rm L}_{Z,1}$), the upper bound on the phase-error rate ($e^{(\rm ph)\hspace{.05cm}\rm U}_{1}$) and the error rate in the key basis ($e_{Z}$). In the simulations, the error correction leakage is modeled as $\lambda_{\rm EC}=1.16M^{Z}_{\alpha}h(e_{Z})$, meaning that we assume an error-correction efficiency of 1.16 and a perfect knowledge of the bit-error rate for simplicity. In this regard, we remark that the more sensitive approach of setting a pre-fixed threshold bit-error rate to assure the robustness of the EC would only have a negligible impact on the SKR, due to the fairly large numbers of signals transmitted in the experiments. In addition, this issue is independent of the correctness of the protocol, which can be guaranteed by simply performing an error verification step after EC.

The experimental results are greatly consistent with our simulation results. In particular, we note that the experiment with $N=10^9$ signals outperforms the simulated SKR to some extent. This is so because, compared to the simulated observables, the set of actually observed key-basis measure counts leads to a slightly tighter confidence interval for the number of single-photon counts $M_{Z,1}$.
\begin{table}[htbp]
\caption{\label{tab: experimental results} Results of the experiments}
\begin{tabular}{cc|cccc|c}
\hline
$\mathrm{f.\hspace{.1cm}loss}$ & $N$ & $M^{Z}_{\alpha}$ & $M^{\rm L}_{Z,1}$ & $e^{(\rm ph)\hspace{.05cm}\rm U}_{1}$ & $e_{Z}$ & $K$ \\
\hline
\hspace{.15cm}6 dB & $10^{10}$ & $1.4\times{}10^{7}$  & $1.0\times{}10^{7}$ & $6.2\%$  & $3.4\%$ & $3.3\times{}10^{-4}$ \\
10 dB & $10^{10}$ & $4.0\times{}10^{6}$ & $3.2\times{}10^{6}$ & $7.1\%$  & $3.4\%$ & $9.9\times{}10^{-5}$ \\
\hspace{.15cm}6 dB & $10^{9}$ & $1.0\times{}10^{6}$ & $7.7\times{}10^{5}$ & $7.6\%$  & $3.4\%$ & $2.1\times{}10^{-4}$ \\
\hline
\end{tabular}
\end{table}

\hfill

In summary, we have experimentally demonstrated FP-QKD exploiting very recent theoretical achievements, and solving the difficulty of designing a stable FP transmitter and an efficient local post-selection system. Our single-laser structure allows to avoid wavelength side-channels, and our time-bin encoding approach benefits from a high stability in the fiber and a simpler experimental layout. What is more, our post-selection method can accurately select the required states and ignore the undesired rounds to reduce the throughput. On top of it, we have assessed the finite-key scenario, which implies that our results are of immediate practical relevance.

In view of the fact that FP-QKD eliminates all modulator side-channels, our work promotes the practical security of QKD and might play a central role in its way to standardization.

(\textit{Note added}. Recently, we have become aware of a related work \cite{hu2023proof}.)


\begin{acknowledgments}
This work has been supported by the National Key Research And Development Program of China (Grant No. 2018YFA0306400), the National Natural Science Foundation of China ( No. 62271463, 62171424, and 62105318), the China Postdoctoral Science Foundation (2022M723064,2021M693098); and the Anhui Initiative in Quantum Information Technologies. MC and VZ acknowledge support from Cisco Systems Inc., the Galician Regional Government (consolidation of Research Units: AtlantTIC), the Spanish Ministry of Economy and Competitiveness (MINECO), the Fondo Europeo de Desarrollo Regional (FEDER) through the grant No. PID2020-118178RB-C21, MICIN with funding from the European Union NextGenerationEU (PRTR-C17.I1) and the Galician Regional Government with own funding through the “Planes Complementarios de I+D+I con las Comunidades Autónomas” in Quantum Communication, the European Union’s Horizon Europe Framework Programme under the Marie Sklodowska-Curie Grant No. 101072637 (Project QSI) and the project “Quantum Security Networks Partnership” (QSNP, grant agreement No 101114043).
\end{acknowledgments}

\begin{widetext}
\appendix*
\section*{Supplemental Material I: post-selecting the decoy-states and encoding}

\begin{figure}[htbp]
	\includegraphics[width=11cm]{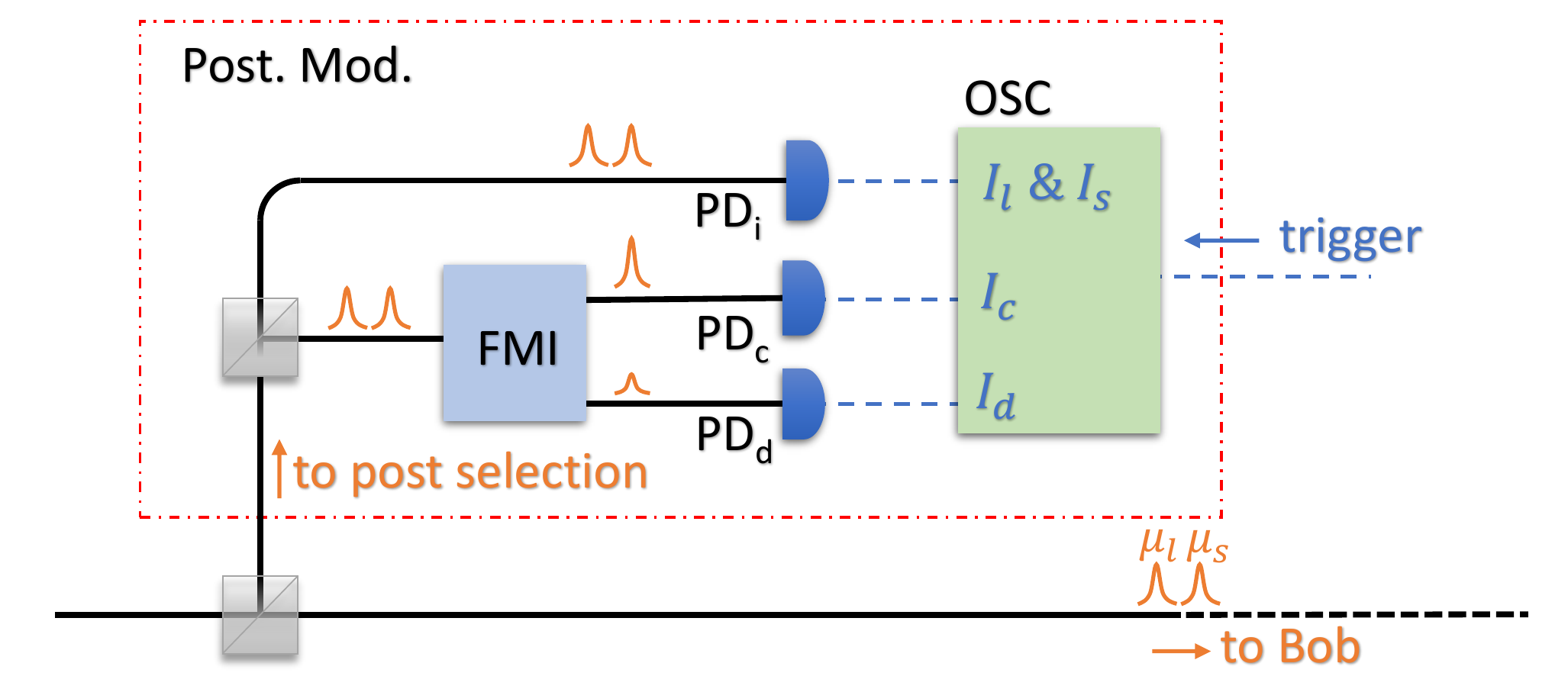}
	\caption{\label{fig: post-selection module} Diagram illustrating the post-selection module, which is marked in the figure with a dash-dotted red line box. Post. Mod.: post-selection module; FMI: Faraday-Michelson interferometer; OSC: oscilloscope; $\rm{PD_i}$: photon diode to measure the intensities of the early and late time-bin pulses, $\rm{PD_c}$($\rm{PD_d}$): photon diode for measuring the constructive (destructive) interference of the FMI; black solid lines: fibers; dashed blue lines: cables.
  }
\end{figure}

Here we provide all the details of the post-selection. As shown in Fig.~\ref{fig: post-selection module}, Alice splits each generated pulse into two pulses. One is sent to Bob to through the quantum channel and the other one is sent to her post-selection module. In this module, Alice measures the random states with three high-speed PDs. Precisely, $\rm{PD_i}$ is employed to measure the intensities of both the early and the late time-bins of each pulse, while $\rm{PD_c}$ ($\rm{PD_d}$) is employed to measure the constructive (destructive) interference of the FMI. The high-speed PDs transfer the light pulses to electronic signals and the OSC records the latter whenever it receives a trigger signal from Bob corresponding to a successful event. According to the OSC's records, Alice can calculate the intensity of the generated pulses based on the outcomes of the three PDs \cite{roberts2018patterning,kang2023patterning}. In particular, we denote the intensities of the early time-bin, the late time-bin, the constructive interference and the destructive interference by $\mu_e$, $\mu_l$, $\mu_c$, and $\mu_d$, respectively. Defining
\begin{equation}\label{mu}
  \begin{aligned}
    \ket{\sqrt{\mu}\cos(\theta/2)} \ket{\sqrt{\mu}\sin(\theta/2) e^{i\phi}}, 
  \end{aligned}
\end{equation}
as Alice's random coherent states representing the two time-bins, the associated single-photon state is
\begin{equation}\label{Bloch_sphere}
  \begin{aligned}
    \cos(\theta/2) \ket{e} + e^{i\phi} \sin(\theta/2) \ket{l} 
  \end{aligned}
\end{equation}
where $\mu$ in Eq.~(\ref{mu}) is the intensity of the coherent states, $\ket{e}$ and $\ket{l}$ respectively denote the early and the late time-bins, and $\theta$ ($\phi$) represents the polar (azimuthal) angle on the Bloch sphere. The quantities $\mu$, $\theta$, and $\phi$ are the necessary parameters to depict Alice's random coherent states, and they can be calculated using the observables $\mu_e$, $\mu_l$, $\mu_c$, and $\mu_d$.
Noting that $\sqrt{\mu_e} = \sqrt{\mu}\cos(\theta/2)$ and $\sqrt{\mu_l} = \sqrt{\mu}\sin(\theta/2)$, we have that $\mu$ is given by
\begin{equation}
  \begin{aligned}
    \mu = \mu_e + \mu_l, 
  \end{aligned}
\end{equation}
while $\theta$ can be obtained as
\begin{equation}
  \begin{aligned}
    \theta = 2\arccos (\sqrt{\frac{\mu_e}{\mu_e + \mu_l}}).
  \end{aligned}
\end{equation}
The angle $\phi$ describes the relative phase between the early time-bin and the late time-bin, and it satisfies
\begin{equation}
  \begin{aligned}
    &\mu_e + \mu_l + 2\sqrt{\mu_e\mu_l} \cos\phi = \mu_c, \\
    &\mu_e + \mu_l - 2\sqrt{\mu_e\mu_l} \cos\phi = \mu_d, \\
    &\mu_e + \mu_l = \mu_c + \mu_d = \mu,
  \end{aligned}
\end{equation}
implying that
\begin{equation}
  \begin{aligned}
    |\phi| = \arccos\left( \frac{\mu_c - \mu_d}{ 4\sqrt{\mu_e\mu_l} }   \right).
  \end{aligned}
\end{equation}

The above parameters ($\mu$, $\theta$, and $\phi$) determine the post-selected basis, key bit, and decoy intensity interval, and as illustrated in Fig.~2~(a) of the main text, Alice pre-decides the threshold values $\Delta\theta_Z$, $\Delta\theta_X$ and $\Delta\phi_X$ that delimitate the acceptance regions. When $\theta < \Delta\theta_Z$ ($\theta >\pi - \Delta\theta_Z$), Alice records the $Z$ basis with raw key bit 0 (1). When $\left\lvert \theta - \pi/2 \right\rvert < \Delta \theta_X $ Alice records the $X$ basis, $\left\lvert \phi  \right\rvert < \Delta\phi_X $ ($\left\lvert \phi-\pi  \right\rvert < \Delta\phi_X $) determining the bit value 0 (1). When none of the above conditions is satisfied, the signal is discarded.

Under the condition that both the basis and the bit value have been determined, Alice post-selects her decoy intensity. The decoy-state post-selection is illustrated in Fig.~2~(b) of the main text, where $\Delta Z$, $\Delta X$, $t_1$, and $t_2$ are pre-defined threshold values. In particular, the quantities $\Delta Z$ and $\Delta X$ satisfy
\begin{equation}
  \begin{aligned}
    &\tan(\Delta Z) = \tan^2( \Delta\theta_Z/2) ,\\
    &\tan(\pi/4 -\Delta X/2) = \tan^2(\pi/4 -\Delta \theta_X/2),\\
  \end{aligned}
\end{equation}
which indicates that
\begin{equation}
  \begin{aligned}
    & \Delta Z = \arctan\left[ \frac{\sin^2(\Delta \theta_Z)}{\left(1+ \cos(\Delta \theta_Z)\right)^2 }   \right], \\
    & \Delta X = \frac{\pi}{2} - 2\arctan\left[ \frac{\cos^2(\Delta \theta_X)}{\left(1+ \sin( \Delta\theta_X)\right)^2 }   \right]  .
  \end{aligned}
\end{equation}
According to the observables $\mu_e$ and $\mu_l$, and also to the threshold values $t_1$ and $t_2$, Alice records the post-selected intensity interval. To be precise, in terms of the maximum intensity, $\mu_{\rm{max}}$, of each time-bin (either early or late), the decoy intervals are delimited by $\mu<t_2\mu_{\rm{max}}$ (\textit{i.e.} $\mu \in I_{\gamma}$, as shown in Fig.~2~(b) of the main text), $\mu<t_1\mu_{\rm{max}}$ (\textit{i.e.} $\mu \in I_{\beta}$) and $\mu<2\mu_{\rm{max}}$ (\textit{i.e.} $\mu \in I_{\alpha}$). Note that this last interval corresponds to no intensity restriction at all, because the maximum intensity is $2\mu_{\rm{max}}$ (corresponding to both the early and the late time bins having an intensity of $\mu_{\rm{max}}$). In summary, the post-selection regions (in terms of $\theta$, $\phi$ and $\mu$) are defined as 
\begin{equation}\label{key_regions}
\Omega^{Z}_{0,j}=\biggl\{\phi\in(-\pi,\pi],\theta\in\left(0,\Delta\theta_{Z}\right),\mu\in{}I_{j}\biggr\},\hspace{.2cm}\Omega^{Z}_{1,j}=\biggl\{\phi\in(-\pi,\pi],\theta\in\left(\pi-\Delta\theta_{Z},\pi\right),\mu\in{}I_{j}\biggr\}
\end{equation}
for the key basis, and
\begin{eqnarray}\label{test_regions}
&&\Omega^{X}_{0,j}=\biggl\{\phi\in\left(-\Delta\phi_{X},\Delta\phi_{X}\right),\theta\in\left(\frac{\pi}{2}-\Delta\theta_{X},\frac{\pi}{2}+\Delta\theta_{X}\right),\mu\in{}I_{j}\biggr\}\nonumber \\
&&\Omega^{X}_{1,j}=\biggl\{\phi\in\left(\pi-\Delta\phi_{X},\pi+\Delta\phi_{X}\right),\theta\in\left(\frac{\pi}{2}-\Delta\theta_{X},\frac{\pi}{2}+\Delta\theta_{X}\right),\mu\in{}I_{j}\biggr\}
\end{eqnarray}
for the test basis, where $j\in\{\alpha,\beta,\gamma\}=:\Gamma$. In addition, for later convenience we define $\Omega^{Z}_{j}=\Omega^{Z}_{0,j}\cup\Omega^{Z}_{1,j}$ and $\Omega^{X}_{j}=\Omega^{X}_{0,j}\cup\Omega^{X}_{1,j}$.

~







\section*{Supplemental Material II: Secret key rate}

In what follows we include all the necessary formulas to calculate the secret key rate of our experiments. Note that the security analysis below follows the methods presented in~\cite{Zapatero2023finitekey}, and we refer the reader to this work (and also to~\cite{zapatero2023fully}) for detailed derivations.

\subsection*{Characterization of the passive transmitter}
The output probability density function of the passive transmitter, to be understood as the probability density function of generating a phase-randomized weak coherent pulse with intensity $\mu$ and polarization $(\theta,\phi)$ in the Bloch sphere spanned by $\ket{e}$ and $\ket{l}$ ---see Eq.~(\ref{Bloch_sphere})---, factors as
\begin{equation}\label{PDF_3}
f_{\boldsymbol{\phi},\boldsymbol{\theta},\boldsymbol{\mu}}(\phi,\theta,\mu)=f_{\boldsymbol{\phi}}(\phi)\times{}f_{\boldsymbol{\theta},\boldsymbol{\mu}}(\theta,\mu),
\end{equation}
where
\begin{equation}\label{PDF_2}
f_{\boldsymbol{\phi}}(\phi)=\frac{1}{2\pi}\hspace{.2cm}\mathrm{and}\hspace{.2cm}f_{\boldsymbol{\theta},\boldsymbol{\mu}}(\theta,\mu)=\frac{1}{\mu_{\mathrm{max}}{}\pi^{2}\sqrt{1-\displaystyle{\frac{\mu}{\mu_{\mathrm{max}}}\cos^{2}\left(\frac{\theta}{2}\right)}}\sqrt{1-\displaystyle{\frac{\mu}{\mu_{\mathrm{max}}}\sin^{2}\left(\frac{\theta}{2}\right)}}}
\end{equation}
for $\phi\in(-\pi,\pi]$, $\theta\in[0,\pi]$ and $\mu\in\left[0,\mu_{\mathrm{max},\theta}\right)$, $\mu_{\mathrm{max},\theta}$ being defined as
\begin{equation}
\mu_{\mathrm{max},\theta}=\min\left\{\frac{\mu_{\rm{max}}}{\cos^{2}\left(\theta/2\right)},\frac{\mu_{\rm{max}}}{\sin^{2}\left(\theta/2\right)}\right\}.
\end{equation}
We recall that $\mu_{\rm{max}}$ (determined experimentally) denotes the highest possible intensity in each time-bin. 

Below we shall use the notation $\left\langle\cdot\right\rangle_{\Omega}$ to denote the triple integral of any input ``$\cdot$" weighted by $f_{\boldsymbol{\phi},\boldsymbol{\theta},\boldsymbol{\mu}}(\phi,\theta,\mu)$ in the region $\Omega$ of the $(\phi,\theta,\mu)$-space~\cite{zapatero2023fully}.

 \subsection*{Parameter estimation}\label{key_rate}
 The observables of the protocol are the key-basis numbers of counts $\{M_{j}^{Z}\}_{j\in\Gamma}$, the test-basis numbers of counts $\{M_{j}^{X}\}_{j\in\Gamma}$ and the test-basis numbers of error counts
 $\{m_{j}^{X}\}_{j\in\Gamma}$, where $\Gamma$ denotes again the set of intensity intervals.
 
According to the finite key analysis of~\cite{Zapatero2023finitekey}, an $(\sqrt{\epsilon_{\rm PE}}+\epsilon_{\rm PA}+\delta)$-secret key of length

\begin{equation}\label{keylength}
l=\left\lfloor{M^{\rm L}_{Z,1}\left[1-h\left(e^{(\rm ph)\hspace{.05cm}\rm U}_{1}\right)\right]-\lambda_{\rm EC}-\log\left(\frac{1}{2\epsilon_{\rm cor}\epsilon_{\rm PA}^{2}\delta}\right)}\right\rfloor
\end{equation}
can be extracted via privacy amplification, where $\epsilon_{\rm PE}$ ($\epsilon_{\rm PA}$) is an upper bound on the error probability of the parameter estimation (privacy amplification) and $\epsilon_{\rm cor}$ is an upper bound on the correctness parameter. Similarly, $\delta$ denotes an extra smoothing parameter of the analysis. On the other hand, $M^{\rm L}_{Z,1}$ is a lower bound on the number of key-basis single-photon counts ($M_{Z,1}$), $e^{(\rm ph)\hspace{.05cm}\rm U}_{1}$ is an upper bound on the single-photon phase-error rate of the sifted key data, and $\lambda_{\rm EC}$ denotes the error correction leakage. As usual, $h(x)$ stands for the binary entropy function, and the secret key rate is computed as $K=l/N$, $N$ denoting the number of transmission rounds.

Coming next, we provide a list of formulas to evaluate $M^{\rm L}_{Z,1}$ and $e^{(\rm ph)\hspace{.05cm}\rm U}_{1}$ in terms of the observables of the protocol. For simplicity, a common error probability $\epsilon$ shall be assumed for every usage of a concentration inequality in what follows. As a consequence, careful counting reveals that the error probability of the parameter estimation method presented here is upper-bounded by $\epsilon_{\rm PE}=17\epsilon$, in virtue of the union bound.

Notably as well, we further set $\epsilon_{\rm PA}=\epsilon$, $\delta=\epsilon$ and $\epsilon_{\rm cor}=\epsilon$ for the calculation of the secret key rates in Fig.~4 of the main text. Putting it all together, this leads to an overall secrecy parameter of $\epsilon_{\rm sec}=\sqrt{17\epsilon}+2\epsilon$, and also to a correctness parameter of $\epsilon_{\rm cor}=\epsilon$~\cite{Zapatero2023finitekey}.

\subsubsection*{Calculation of $M_{Z,1}^{\rm L}$}
$M_{Z,1}^{\rm L}$ is computed as~\cite{Zapatero2023finitekey}
 \begin{equation}
 M_{Z,1}^{\rm L}=\bar{K}^{\rm L}_{N,\epsilon}\Bigl(N{}(1-q_{X})\left\langle{e^{-\mu}{\mu}}\right\rangle_{\Omega^{Z}_{\alpha}}y_{1}^{\rm L}\Bigr),
 \end{equation}
 where $\bar{K}^{\rm L}_{N,\epsilon}(x)$ is a reverse Kato function~\cite{kato2020concentration} defined in Eq.~(\ref{bar_k_L}), $q_{X}$ is the test basis probability of Bob's receiver, and $y_{1}^{\rm L}$ is the solution to the linear program below:
 \begin{eqnarray}\label{lp_1}
&&\min\quad y_{\alpha,1}^{Z}\nonumber\\
&&\textup{s.t.}\hspace{.3cm}\displaystyle\sum_{n=0}^{n_{\rm cut}}p^{Z}_{n|j}y^{Z}_{j,n}\leq{}Q_{j}^{Z}\leq\displaystyle\sum_{n=0}^{n_{\rm cut}}p^{Z}_{n|j}y^{Z}_{j,n}+1-\displaystyle\sum_{n=0}^{n_{\rm cut}}p^{Z}_{n|j}\hspace{.2cm}\mathrm{for}\hspace{.2cm}j\in\Gamma,\nonumber\\
&&\hspace{.8cm}\frac{K^{\rm L}_{N,\epsilon}(M_{j}^{Z})}{N{}(1-q_{X})\left\langle{1}\right\rangle_{\Omega_{j}^{Z}}}\leq{Q_{j}^{Z}}\leq{}\frac{K^{\rm U}_{N,\epsilon}(M_{j}^{Z})}{N{}(1-q_{X})\left\langle{1}\right\rangle_{\Omega_{j}^{Z}}}\hspace{.2cm}\mathrm{for}\hspace{.2cm}j\in\Gamma,\nonumber\\
&&\hspace{.8cm}\left\lvert{y^{Z}_{j,n}-y^{Z}_{k,n}}\right\rvert\leq{}D\bigl(\sigma^{Z}_{j,n},\sigma^{Z}_{k,n}\bigr)\hspace{.2cm}\mathrm{for}\hspace{.2cm}j\in\Gamma,k\in\Gamma,\ j\neq{}k,\ n=2,\ldots{},n_{\rm cut},\nonumber\\
&&\hspace{.8cm}y^{Z}_{j,1}=y^{Z}_{k,1}\hspace{.2cm}\mathrm{for}\hspace{.2cm}j\in\Gamma,k\in\Gamma,\ j\neq{}k,\nonumber\\
&&\hspace{.8cm}0\leq{}y^{Z}_{j,n}\leq{}1\hspace{.2cm}\hspace{.2cm}\mathrm{for}\hspace{.2cm}j\in\Gamma,\ n=0,\ldots,n_{\rm cut},
\end{eqnarray}
where $n_{\rm cut}$ is a cutoff photon number for the decoy constraints, $\left\{p^{Z}_{n|j}=\left\langle{e^{-\mu}{\mu}^{n}/{n!}}\right\rangle_{\Omega^{Z}_{j}}/\langle{1}\bigr\rangle_{\Omega^{Z}_{j}}\right\}_{j,n}$ are the key-basis photon-number statistics, $\left\{y^{Z}_{j,n}=\sum_{u}y^{Z\hspace{.05cm}(u)}_{j,n}/N\right\}_{j,n}$ are the key-basis conditional yields averaged over protocol rounds, with $y^{Z\hspace{.05cm}(u)}_{j,n}=p^{(u)}\bigl(\mathrm{click}|E_{u-1},\sigma^{Z}_{j,n},Z\bigr)$ denoting the detection probability in round $u$ conditioned on the state of Eve's classical register up to that round, $E_{u-1}$~\cite{Zapatero2023finitekey}, and on the joint event that Bob measures in the key basis and a key-basis Fock state with $n$ photons and setting $j$ is post-selected,
\begin{equation}\label{Z_state}
\sigma^{Z}_{j,n}=\frac{\left\langle{e^{-\mu}{\mu}^{n}/n!\ketbra{n}{n}_{\theta,\phi}}\right\rangle_{\Omega^{Z}_{j}}}{\left\langle{e^{-\mu}{\mu}^{n}/n!}\right\rangle_{\Omega^{Z}_{j}}}.
\end{equation}
On the other hand, $Q^{Z}_{j}=\sum_{u}Q^{Z\hspace{.05cm}(u)}_{j}/N$ are the key-basis conditional gains averaged over protocol rounds, for $Q^{Z\hspace{.05cm}(u)}_{j}=\sum_{n=0}^{\infty}p^{Z}_{n|j}y^{Z\hspace{.05cm}(u)}_{j,n}$, the quantities $K^{\rm L}_{N,\epsilon}(x)$ and $K^{\rm U}_{N,\epsilon}(x)$ are the Kato functions respectively defined in Eq.~(\ref{k_L}) and Eq.~(\ref{k_U}), and
\begin{equation}
D(\rho,\tau)=\frac{1}{2}\Tr\left[\sqrt{\left(\rho-\tau\right)^{2}}\right]
\end{equation}
stands for the trace distance (TD) between $\rho$ and $\tau$. A brief discussion on how to compute the necessary TD values is provided at the end of this section.

For the obtention of the results presented in Fig.~4 of the main text, the threshold photon-number in the linear programs was set to $n_{\rm cut}=4$, and all the numerical calculations and simulations were performed with Matlab.

\subsubsection*{Calculation of $e^{(\rm ph)\hspace{.05cm}\rm U}_{1}$}
The parameter $e^{(\rm ph)\hspace{.05cm}\rm U}_{1}$ is calculated as
\begin{equation}
e^{(\rm ph)\hspace{.05cm}\rm U}_{1}=\min\left\{\displaystyle{\frac{1}{2},\frac{m^{(\rm ph)\hspace{.05cm}\rm U}_{Z,1}}{M_{Z,1}^{\rm L}}}\right\},
\end{equation}
where $m^{(\rm ph)\hspace{.05cm}\rm U}_{Z,1}$ denotes an upper bound on the number of single-photon phase errors~\cite{Zapatero2023finitekey} in the sifted key data. This bound is given by
\begin{equation}
m^{(\rm ph)\hspace{.05cm}\rm U}_{Z,1}=\frac{M_{Z,1}^{\rm U}}{M_{X,1}^{\mathrm{ideal,L}}}m_{X,1}^{\mathrm{ideal,U}}+\Upsilon(M_{Z,1}^{\rm U},M_{X,1}^{\mathrm{ideal,L}},\epsilon),
\end{equation}
where $\Upsilon(x,y,z)=\sqrt{(x+y)x(y+1)\ln\left(z^{-1}\right)/2y^2}$ is a deviation term arising from Serfling's inequality~\cite{serfling1974probability}, the parameter
\begin{equation}
M_{Z,1}^{\rm U}=\bar{K}^{\rm U}_{N,\epsilon}\Bigl(N{}(1-q_{X})\left\langle{e^{-\mu}{\mu}}\right\rangle_{\Omega^{Z}_{\alpha}}y_{1}^{\rm U}\Bigr)
\end{equation}
is an upper bound on $M_{Z,1}$ ---$\bar{K}^{\rm U}_{N,\epsilon}$ being the reverse Kato function defined in Eq.~(\ref{bar_k_U}), and $y_{1}^{\rm U}$ being the solution to the linear program of Eq.~(\ref{lp_1}) but under a maximization condition---, the quantity
\begin{equation}
M_{X,1}^{\mathrm{ideal,L}}=\bar{K}^{\rm L}_{N,\epsilon}\Bigl(N{}q_{X}\left\langle{e^{-\mu}{\mu}}\right\rangle_{\Omega^{X}_{\alpha}}\lambda_{X}y_{1}^{\rm L}\Bigr)
\end{equation}
is a lower bound on the number $M_{X,1}^{\mathrm{ideal}}$ of test-basis single-photon counts triggered by perfectly prepared states ---$\lambda_{X}=\langle{e^{-\mu}{\mu}\sin{\theta}\cos{\phi}}\rangle_{\Omega^{X}_{0,\alpha}}\bigl/\langle{e^{-\mu}{\mu}}\rangle_{\Omega^{X}_{0,\alpha}}$ denoting the expected ratio of perfect $\ketbra{+}{+}$ states among all single-photon states prepared in the acceptance region $\Omega^{X}_{0,\alpha}$---, and
\begin{equation}
m_{X,1}^{\mathrm{ideal,U}}=\bar{K}^{\rm U}_{N,\epsilon}\Bigl(N{}q_{X}\left\langle{e^{-\mu}{\mu}}\right\rangle_{\Omega^{X}_{\alpha}}\lambda_{X}e_{X,1}^{\mathrm{ideal,U}}\Bigr)
\end{equation}
is an upper bound on the number $m_{X,1}^{\mathrm{ideal}}$ of test-basis single-photon error counts triggered by perfectly prepared states, with $e_{X,1}^{\mathrm{ideal,U}}$ denoting the solution to the following linear program
 \begin{eqnarray}\label{lp_2}
&&\max\quad e_{X,1}^{\rm ideal}\nonumber\\
&&\textup{s.t.}\hspace{.3cm}\displaystyle\sum_{n=0}^{n_{\rm cut}}p^{X}_{n|j}e_{j,n}^{X}\leq{}E^{X}_{j}\leq\displaystyle\sum_{n=0}^{n_{\rm cut}}p^{X}_{n|j}e_{j,n}^{X}+1-\displaystyle\sum_{n=0}^{n_{\rm cut}}p^{X}_{n|j}\hspace{.2cm}\mathrm{for}\hspace{.2cm}j\in\Gamma,\nonumber\\
&&\hspace{.8cm}\frac{K^{\rm L}_{N,\epsilon}(m_{j}^{X})}{N{}q_{X}\left\langle{1}\right\rangle_{\Omega_{j}^{X}}}\leq{E^{X}_{j}}\leq{}\frac{K^{\rm U}_{N,\epsilon}(m_{j}^{X})}{N{}q_{X}\left\langle{1}\right\rangle_{\Omega_{j}^{X}}}\hspace{.2cm}\mathrm{for}\hspace{.2cm}j\in\Gamma,\nonumber\\
&&\hspace{.8cm}\left\lvert{e_{j,n}^{X}-e_{k,n}^{X}}\right\rvert\leq{}D\bigl(\sigma_{j,n}^{+},\sigma_{k,n}^{+}\bigr)\hspace{.2cm}\mathrm{for}\hspace{.2cm}j\in\Gamma,\ j\neq{}k,\ n=1,\ldots{},n_{\rm cut},\nonumber\\
&&\hspace{.8cm}\lambda_{j}e_{X,1}^{\mathrm{ideal}}\leq{}e_{j,1}^{X}-(1-\lambda_{j})y_{1}^{\rm L}\frac{1}{2}\hspace{.2cm}\mathrm{for}\hspace{.2cm}j\in\Gamma,\nonumber\\
&&\hspace{.8cm}0\leq{}e_{j,n}^{X}\leq{}1\hspace{.2cm}\hspace{.2cm}\mathrm{for}\hspace{.2cm}j\in\Gamma,\ n=0,\ldots,n_{\rm cut}.
\end{eqnarray}
Here, $\left\{p^{X}_{n|j}=\left\langle{e^{-\mu}{\mu}^{n}/{n!}}\right\rangle_{\Omega^{X}_{j}}/\langle{1}\bigr\rangle_{\Omega^{X}_{j}}\right\}_{j,n}$ stand for the test-basis photon-number statistics, $\left\{e^{X}_{j,n}=\sum_{u}e^{X\hspace{.05cm}(u)}_{j,n}/N\right\}_{j,n}$ stand for the test-basis conditional error yields averaged over protocol rounds (subject to the random assignment of multiple clicks for the purpose of having ``+" and ``-" as the only two possible test-basis measurement outcomes), and $e^{X\hspace{.05cm}(u)}_{j,n}=\left[p^{(u)}\bigl(-|E_{u-1},\sigma^{+}_{j,n},X\bigr)+p^{(u)}\bigl(+|E_{u-1},\sigma^{-}_{j,n},X\bigr)\right]/2$, where
\begin{equation}\label{+_state}
\sigma^{+}_{j,n}=\frac{\left\langle{e^{-\mu}{\mu}^{n}/n!\ketbra{n}{n}_{\theta,\phi}}\right\rangle_{\Omega^{X}_{0,j}}}{\left\langle{e^{-\mu}{\mu}^{n}/n!}\right\rangle_{\Omega^{X}_{0,j}}}
\end{equation}
and
\begin{equation}\label{-_state}
\sigma^{-}_{j,n}=\frac{\left\langle{e^{-\mu}{\mu}^{n}/n!\ketbra{n}{n}_{\theta,\phi}}\right\rangle_{\Omega^{X}_{1,j}}}{\left\langle{e^{-\mu}{\mu}^{n}/n!}\right\rangle_{\Omega^{X}_{1,j}}}
\end{equation}
denote the test-basis Fock states. Finally, $E^{X}_{j}=\sum_{u}E^{X\hspace{.05cm}(u)}_{j}/N$ are the test-basis conditional error gains averaged over protocol rounds, such that $E^{X\hspace{.05cm}(u)}_{j}=\sum_{n=0}^{\infty}p^{X}_{n|j}e^{X\hspace{.05cm}(u)}_{j,n}$, and $e_{X,1}^{\mathrm{ideal}}=\sum_{u}e_{X,1}^{\mathrm{ideal}\hspace{.05cm}(u)}/N$ for
\begin{equation}
e_{X,1}^{\mathrm{ideal}\hspace{.05cm}(u)}=\frac{p^{(u)}\left(-|E_{u-1},\ketbra{+}{+},X\right)+p^{(u)}\left(+|E_{u-1},\ketbra{-}{-},X\right)}{2}.
\end{equation}
Namely, $e_{X,1}^{\mathrm{ideal}\hspace{.05cm}(u)}$ is the contribution to $e^{X\hspace{.05cm}(u)}_{j,1}$ arising from the ideal single-photon states $\ket{+}$ and $\ket{-}$.

\subsection*{Kato's inequality confidence intervals}
Kato's inequality is a concentration inequality for the sum of dependent random variables. The Kato functions that provide the confidence intervals required for our finite-key analysis are defined in what follows, where we distinguish between \textit{direct functions} ---which provide bounds on the sum of conditional expectations--- and \textit{reverse functions} ---which provide bounds on the actual realization of the sum---.
The direct Kato functions are presented first. On the one hand,
\begin{equation}\label{k_L}
K^{\rm L}_{N,\epsilon}\left(x\right)=x-\left[b+a\left(2x/N-1\right)\right]\sqrt{N}
\end{equation}
for
\begin{eqnarray}
&&a={{3\left\{9\sqrt{2}N\left(N-2\tilde{x}\right)\sqrt{-\ln{\epsilon}\left[9\tilde{x}\left(N-\tilde{x}\right)-2N\ln{\epsilon}\right]}+16N^{3/2}\ln^{2}{\epsilon}-72\tilde{x}\sqrt{N}\left(N-\tilde{x}\right)\ln {\epsilon}\right\}}\over{4\left(9N-8\ln {\epsilon}\right)\left[9\tilde{x}\left(N-\tilde{x}\right)-2N\ln {\epsilon}\right]}},\nonumber \\
&&b={{\sqrt{18Na^2-\left(16a^2-24\sqrt{N}a+9N\right)\ln{\epsilon}}}\over{3\sqrt{2N}}},
\end{eqnarray}
where $\tilde{x}$ is an a priori guess of $x$. Importantly, in the obtention of the results presented in Fig.~4 of the main text, all the necessary guesses for Kato's inequality were selected under the assumption of a perfect characterization of the experimental setup. Note, however, that this assumption can be easily removed by carefully modelling the channel or by using the outcomes of previous experiments.%

On the other hand,
\begin{equation}\label{k_U}
K^{\rm U}_{N,\epsilon}\left(x\right)=x+\left[b+a\left(2x/N-1\right)\right]\sqrt{N}
\end{equation}
for
\begin{eqnarray}
&&a={{3\left\{9\sqrt{2}N\left(N-2\tilde{x}\right)\sqrt{-\ln{\epsilon}\left[9\tilde{x}\left(N-\tilde{x}\right)-2N\ln{\epsilon}\right]}-16N^{3/2}\ln^{2}{\epsilon}+72\tilde{x}\sqrt{N}\left(N-\tilde{x}\right)\ln {\epsilon}\right\}}\over{4\left(9N-8\ln {\epsilon}\right)\left[9\tilde{x}\left(N-\tilde{x}\right)-2N\ln{\epsilon}\right]}},\nonumber \\
&&b={{\sqrt{18Na^2-\left(16a^2+24\sqrt{N}a+9N\right)\ln{\epsilon}}}\over{3\sqrt{2N}}}.
\end{eqnarray}
As for the reverse Kato functions, we have
\begin{equation}\label{bar_k_L}
\bar{K}^{\rm L}_{N,\epsilon}\left(x\right)=\left[\sqrt{N}x+N(a-b)\right]/(2a+\sqrt{N})
\end{equation}
for $a=\max\left\{a',-\sqrt{N}/2\right\}$ with
\begin{equation}\label{ab_L}
a'=\frac{3\sqrt{N}\left\{9\left(N-2\tilde{x}\right)\sqrt{N\ln{\epsilon}\left[N\ln{\epsilon}-18\tilde{x}\left(N-\tilde{x}\right)\right]}-4N\ln^{2}{\epsilon}-9\left(8\tilde{x}^2-8N\tilde{x}+3N^2\right)\ln {\epsilon}\right\}}{4\left\{4N\ln^{2}{\epsilon}+36\left(2\tilde{x}^2-2N\tilde{x}+N^2\right)\ln {\epsilon}+81N\tilde{x}\left(N-\tilde{x}\right)\right\}}
\end{equation}
and
\begin{equation}
b=\frac{1}{3}\sqrt{9a^2-{{\left(4a+3\sqrt{N}\right)^2\ln{\epsilon}}\over{2N}}},
\end{equation}
where, again, $\tilde{x}$ denotes an a priori guess of $x$. Similarly,
\begin{equation}\label{bar_k_U}
\bar{K}^{\rm U}_{N,\epsilon}\left(x\right)=\left[\sqrt{N}x-N(a-b)\right]/(\sqrt{N}-2a)
\end{equation}
for $a=\min\left\{a',\sqrt{N}/2\right\}$ with
\begin{equation}
a'={{3\sqrt{N}\left\{9\left(N-2\tilde{x}\right)\sqrt{N\ln{\epsilon}\left[N\ln{\epsilon}+18\tilde{x}\left(\tilde{x}-N\right)\right]}+4N\ln^{2}{\epsilon}+9\left(8\tilde{x}^2-8N\tilde{x}+3N^2\right)\ln{\epsilon}\right\}}\over{4\left\{4N\ln^{2}{\epsilon}+36\left(2\tilde{x}^2-2N\tilde{x}+N^2\right)\ln{\epsilon}+81N\tilde{x}\left(N-\tilde{x}\right)\right\}}},
\end{equation}
and
\begin{equation}
b={{\sqrt{18Na^2-\left(16a^2-24\sqrt{N}a+9N\right)\ln{\epsilon}}}\over{3\sqrt{2N}}}.
\end{equation}
\subsection{Trace distance values}\label{TD}
From the definition of the TD, it follows that $D(\rho,\tau)=\sum_{i=1}^{d}\abs{\lambda_{i}}$, where the $\lambda_{i}$ are the eigenvalues of $\rho-\tau$. Hence, in order to compute the necessary set of TD values, $\left\{D\bigl(\sigma^{Z}_{j,n},\sigma^{Z}_{k,n}\bigr),\ D\bigl(\sigma_{j,n}^{+},\sigma_{k,n}^{+}\bigr)\right\}_{j,k,n}$, we provide a matrix representation of the density operators involved and numerically diagonalize them. Specifically, let
\begin{equation}
\mathcal{B}_{n}=\left\{\ket{n-k,k}=\frac{a^{\dagger n-k}_{s}a^{\dagger k}_{l}}{\sqrt{(n-k)!k!}}\ket{\rm vac},\ k=0,\ldots,n\right\},
\end{equation}
where $a^{\dagger}_{s}$ ($a^{\dagger}_{l}$) is the creation operator attached to the early (late) time-bin mode. $\mathcal{B}_{n}$ is an orthonormal basis of the Hilbert space of $n$ indistinguishable photons distributed across two modes. Making use of the canonical isomorphism
\begin{equation}
\ket{n,0}\rightarrow{}[1\hspace{.05cm}0\ldots\hspace{.05cm}0]^{t},\hspace{.1cm}\ket{n-1,1}\rightarrow{}[0\hspace{.05cm}1\ldots\hspace{.05cm}0]^{t},\hspace{.05cm}\ldots\hspace{.1cm},\hspace{.1cm}\ket{0,n}\rightarrow{}[0\hspace{.05cm}\ldots\hspace{.05cm}0\hspace{.05cm}1]^{t},
\end{equation}
one can easily obtain density matrices for the Fock states $\sigma^{Z}_{j,n}$, $\sigma^{+}_{j,n}$ and $\sigma^{-}_{j,n}$ ---respectively given in Eq.~(\ref{Z_state}), Eq.~(\ref{+_state}) and Eq.~(\ref{-_state})--- by exploiting the relation
\begin{equation}
\ket{n}_{\theta,\phi}=\frac{1}{\sqrt{n!}}\left[\cos\left(\frac{\theta}{2}\right)a^{\dagger}_{s}+e^{i\phi}\sin\left(\frac{\theta}{2}\right)a^{\dagger}_{l}\right]^{n}\ket{\rm vac}.
\end{equation}

Particularly, the $(r,s)$-th entry of, say $\sigma^{Z}_{j,n}$, is computed as $\bra{n-r+1,r-1}\sigma^{Z}_{j,n}\ket{n-s+1,s-1}$ for $r,s=1,\ldots,n+1$.

\section*{Supplemental Material III: discussion about some open questions}

Fully passive QKD can help QKD users to avoid the side channels that active modulation may introduce. However, we recall that passive QKD is not source-device-independent, and in particular some practical issues need to be carefully treated.

For instance, the operation of the OS is one of them. In our experiment ---just like in Refs.~\cite{wang2023fully,hu2023proof}---, we employ an OS to eliminate the excess pulses. This OS is opened and closed periodically at a fixed rate, and hence it cannot be considered as an active modulator, nor it encodes any information on the output signals. However, the extinction ratio of a practical OS is limited, and thus imperfectly cancelled pulses may leak information about the neighboring pulses. In this respect, a possible solution might be to cascade various optical switches in order to reduce the extinction ratio to a negligible level. Alternatively, one could attempt to model this source of leakage and account for it in the secret key rate, say, in the line of~\cite{navarrete2022improved} for instance.

Another open question is related to the interference. Since the characterization of the source relies on a perfect interference assumption, interference stability is required. With this in mind, even though in our experiments we employed polarization-maintaining devices and polarization-maintaining fibers in the quantum module, it would be desirable to incorporate the effect of imperfect interference in the characterization of the passive source. This will be addressed in a future work.

To finish with, the theoretical analysis also presumes perfect classical measurements at Alice's local detection unit. Notably though, the effect of imperfect measurements may bring some problems similar in nature to state-preparation flaws \cite{tamaki2014loss,pereira2020quantum,zapatero2021security,yin2013measurement,yoshino2018quantum,lu2021intensity,lu2022unbalanced} in actively modulated QKD systems. Again, a convenient solution would be to theoretically account for imperfect classical measurements in Alice's detection unit.

\end{widetext}

\hfill

\bibliography{citations}

\end{document}